\documentclass[fleqn,twoside]{article}
\usepackage{espcrc2}

\usepackage{graphicx}
\usepackage{amssymb}

\newcommand{\beq}{\begin{equation}}
\newcommand{\eeq}{\end{equation}}
\newcommand{\bqa}{\begin{eqnarray}}
\newcommand{\eqa}{\end{eqnarray}}
\newcommand{\etal}{{\em et al.}}

\newcommand{\AmS}{{\protect\the\textfont2
  A\kern-.1667em\lower.5ex\hbox{M}\kern-.125emS}}

\begin{document}

\thispagestyle{empty}

% declarations for front matter
\title{
\vspace{-1.8cm}
\hfill \rm \null \hfill
 \hbox{\normalsize SNUTP-03-025} \\
\vspace{+1.3cm}
Description and comparison of $\overline{\rm Fat7}$ and HYP fat links}

\author{Sundance O. Bilson-Thompson and Weonjong Lee 
        \address{BK21 Research Division, School of Physics, \\ 
	  Seoul National University, Seoul 151-747 South Korea}}

\begin{abstract}
We study various methods of constructing fat links based upon the HYP
(by Hasenfratz \& Knechtli) and $\overline{\rm Fat7}$ (by W. Lee) algorithms.
We present the minimum plaquette distribution for these fat links. This 
enables us to determine which algorithm is most effective at reducing the 
spread of plaquette values - a strong indicator of improved statistics 
for spectrum and static potential measurements, among other quantities.
  
\vspace{1pc}
\end{abstract}

\maketitle

\section{INTRODUCTION}
Despite its prominent role as a useful tool in the non-perturbative study of
QCD, lattice field theory has long struggled with the difficulties of 
accurately representing a continuum field theory in a discrete spacetime. The 
occurrence of discretization errors has lead to the development of a great many
improvement schemes, some more useful than others. One extremely popular method
of improvement has been the use of ``fat links'', in which each ordinary or 
``thin'' link of the lattice, $U_{\mu}(x)$, is replaced by a linear combinaton 
of the thin link and the links adjacent to it. Two particular fattening 
algorithms will be of interest to us in this report, HYP or hypercubic 
blocking~\cite{hasen}, and $\overline{\rm Fat7}$ blocking (that is, 
SU(3)-projected Fat7 blocking~\cite{wlee:0}). These two algorithms have been 
shown to be perturbatively equivalent at one-loop level~\cite{wlee:0,leesharpe}. 
We will present the results of a preliminary numerical investigation to 
determine whether there are any signifiant non-perturbative differences 
between the two algorithms. These results will point the way to more detailed 
investigations.  
 
\section{HYPERCUBIC BLOCKING}
\begin{figure}
\includegraphics[angle=0,width=15pc]{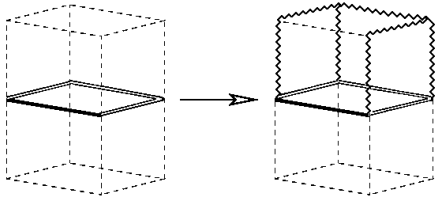}
\caption{HYP fattening. The bold line is the thin link, and the dashed lines 
are the hypercubes attached to it. The double and zig-zig lines are the 
staples at the final and middle levels of fattening respectively. 
The outer level (fourth direction, which must necessarily be constructed first) 
is omitted. Only positively directed staples are shown here.}
\label{fig:HYP}
\end{figure}

The HYP algorithm was proposed as a modification of standard APE 
smearing~\cite{APE}. APE smearing involves constructing a fattened link 
$V_{\mu}(x)$ by adding the sum of staples, weighted by some factor 
$\alpha$, to the thin link,
\beq
V_{\mu}(x)=(1-\alpha)\,U_{\mu}(x) + \frac{\alpha}{6} \sum_{\nu \ne \mu} W_{\nu}(x) 
\label{eq:fatten}
\eeq
followed by projection back to SU(3) (where $W_{\nu}(x)$ refers to 
the sum of three-link staples in the positive and negative 
$\nu$ directions, with central link parallel to $U_{\mu}(x)$). 
Clearly $n$ repeated applications of such an algorithm will 
access the gauge fields at a distance $n$ lattice spacings from 
the original thin link (although smoothing algorithms such as 
cooling and smearing can be viewed as diffusive ``random-walk'' 
processes which primarily affect physics only within a range 
$a\sqrt{n}$, as investigated in~\cite{giacomo}). To localize APE smearing 
within the smallest possible volume, HYP blocking was defined as being 
equivalent to three iterations of APE smearing with the caveat that the staples
at each stage could not be constructed from links which are fattened in the 
same plane as the staple itself. The original thin link is therefore fattened 
by pairs of positively-and-negatively-oriented staples in three directions. 
Each of these staples is constructed from links fattened by staples in only 
two (positive and negative) directions, and each of these staples is 
constructed from links fattened in a single plane (for further details please 
refer to~\cite{hasen}). HYP blocking therefore uses only the staples defined 
by the boundaries of hypercubes attached to the original thin link 
(Fig~\ref{fig:HYP}). Projection back to SU(3) is performed after each level of 
fattening.

We shall henceforth refer to the links fattened by the outermost set of 
staples as the initial-stage fat links, the links fattened by staples formed 
from the initial-stage links as the middle-stage fat links, and the completely 
fattened links (i.e. those fattened by staples formed from middle-stage fat 
links) as the final-stage links. 

Since its introduction, HYP blocking has shown itself to be extremely 
effective at reducing taste-symmetry breaking effects, and has gained 
widespread attention and use. 
  
\section{FAT7 AND $\overline{\rm FAT7}$ BLOCKING}
\label{sec:Fat7}
\begin{figure}
\includegraphics[angle=0,width=15pc]{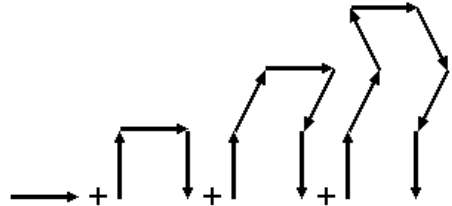}
\caption{The thin link plus 3-link, 5-link, and 7-link staples used to construct the 
Fat7 fat link. In practice the thin link and each of the staples would be weighted 
by an appropriate prefactor.}
\label{fig:fat7}
\end{figure}

Fat7 blocking~\cite{Fat7} is a form of fattening which incorporates not only 
standard three-link staples but also five-link and seven-link staples 
(Fig.~\ref{fig:fat7}). Clearly, with staples up to length seven it is possible 
to traverse all three directions orthogonal to the original thin link, as is 
the case with HYP blocking. However Fat7 blocking as it was originally 
envisaged did not incorporate SU(3) projection. $\overline{\rm Fat7}$ is a 
modification of this algorithm which incorporates SU(3) projection of the final 
sum of link and staples~\cite{wlee:0}.

\begin{figure}
\includegraphics[angle=-90,width=15pc]{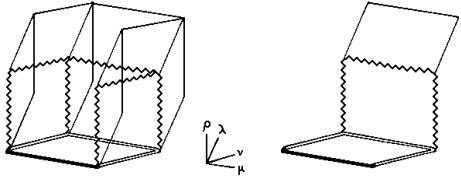}
\caption{The HYP fattening (left) compared with Fat7 (right). The original 
(thin) link is the bold black line, while the double, zig-zig, and solid thin 
lines represent the staples at each successive level of fattening. Only 
positive directions are depicted here, for the sake of clarity.}
\label{fig:HYPvsF7}
\end{figure}

If we construct $\overline{\rm Fat7}$ fat-links in an analogous manner to 
the iterative construction of HYP links we have greater freedom in our 
application of SU(3)-projection. An iterative approach allows us to include 
SU(3) projection at each level of the construction of the link, rather than 
just at the final level. The only difference between HYP and 
$\overline{\rm Fat7}$ is then that in HYP blocking each link in each staple 
is fattened at the subsequent level, whereas in $\overline{\rm Fat7}$ it is 
only the links parallel to the original thin link that are fattened 
(see Fig.~\ref{fig:HYPvsF7}).

We can see that the second covariant derivative operator~\cite{LePage} may be 
used to recursively define a fattened link
\beq
L_{\nu}(\alpha) \cdot U_{\mu}(x) = (1 - 2\alpha)\cdot U_{\mu}(x) + \alpha W_{\nu}(x)
\eeq
($W_{\nu}(x)$ defined as in Eq.~(\ref{eq:fatten})). With the parameter $\alpha$
taking the value one-quarter this operator can be interpreted as suppressing 
flavour-changing gluon interactions by vanishing in the limit as gluon 
momentum approaches the lattice cut-off. The Fat7 link can easily be 
constructed by repeated application of this operator as
\beq
     V_{\mu} = \frac{1}{6}\sum_{{\rm perm}(\nu,\rho,\lambda)} 
        \!\!\!\!\! L_{\nu}(\alpha)\cdot
           \left(L_{\rho}(\alpha)\cdot
	      \left(L_{\lambda}(\alpha)\cdot U_{\mu}\right) \right)
\eeq
where we note explicitly that the sum over permutations of the directions is 
taken. This repeated use of the same operator accounts for the use of a single 
weighting parameter $\alpha$ at each stage, whereas HYP blocking uses three 
parameters $\alpha_1$, $\alpha_2$, $\alpha_3$.

As noted above, to construct the $\overline{\rm Fat7}$ links from the Fat7 
links, SU(3) projection must be performed after the final summation is
taken, but it can also optionally be applied after each or either of the first 
two fattening stages, leaving us with four possible combinations of projection 
schemes to define our $\overline{\rm Fat7}$ link (i.e. projection at the final,
initial+final, middle+final, and initial+middle+final stages). 

\section{COMPARISON OF FATTENING APPROACHES}
Fat links have shown themselves to be extremely useful at overcoming problems 
associated with the lattice formulation. Specifically they have shown a 
reduction in the severity of taste-symmetry breaking associated with staggered 
fermions, and a reduction in the severity of the exceptional configuration 
problem associated with the Wilson fermion formulation. Both of these problems 
can be directly related to short-scale fluctuations of the link values. 
In~\cite{hasen}, the values of the minimum plaquettes calculated on an ensemble
of configurations were used as a probe of the most severe link fluctuations. 
The maximum plaquette values are bounded above, and hence we expect an increase
in the minimum plaquette value of any configuration to indicate a smaller range
of link fluctuations across that configuration, and hence a reduction in the 
scale of the largest short-range link fluctuations.

\begin{figure}
\includegraphics[angle=90,width=15pc]{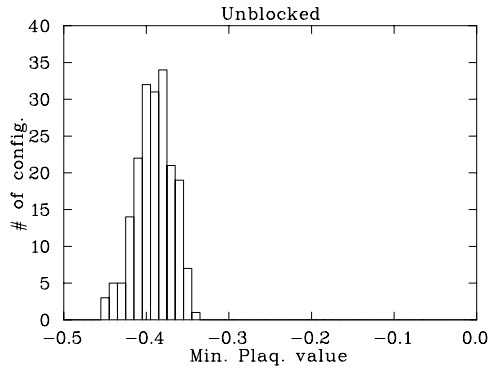}
\caption{Histogram of minimum plaquette values for our ensemble of unblocked 
configurations.}
\label{fig:unblock}
\end{figure}
The process of fattening tends to increase the minimum plaquette values of 
different configurations to differing degrees, spreading the minimum values 
over quite a wide range (as can be seen by comparing Figs.~\ref{fig:unblock} 
and Figs.~\ref{fig:HYP1} and \ref{fig:HYP2}). We expect that certain measurements 
will achieve a lower level of statistical uncertainty if performed on an ensemble of 
configurations having a narrow spread of minimum plaquette values. One of the aims 
of this preliminary study will be to identify whether the $\overline{\rm Fat7}$ 
blocking algorithm is capable of producing a narrower spread of minimum 
plaquette values than HYP, and is hence worthy of a more detailed 
investigation.

\section{RESULTS}
\begin{figure}
\includegraphics[angle=90,width=15pc]{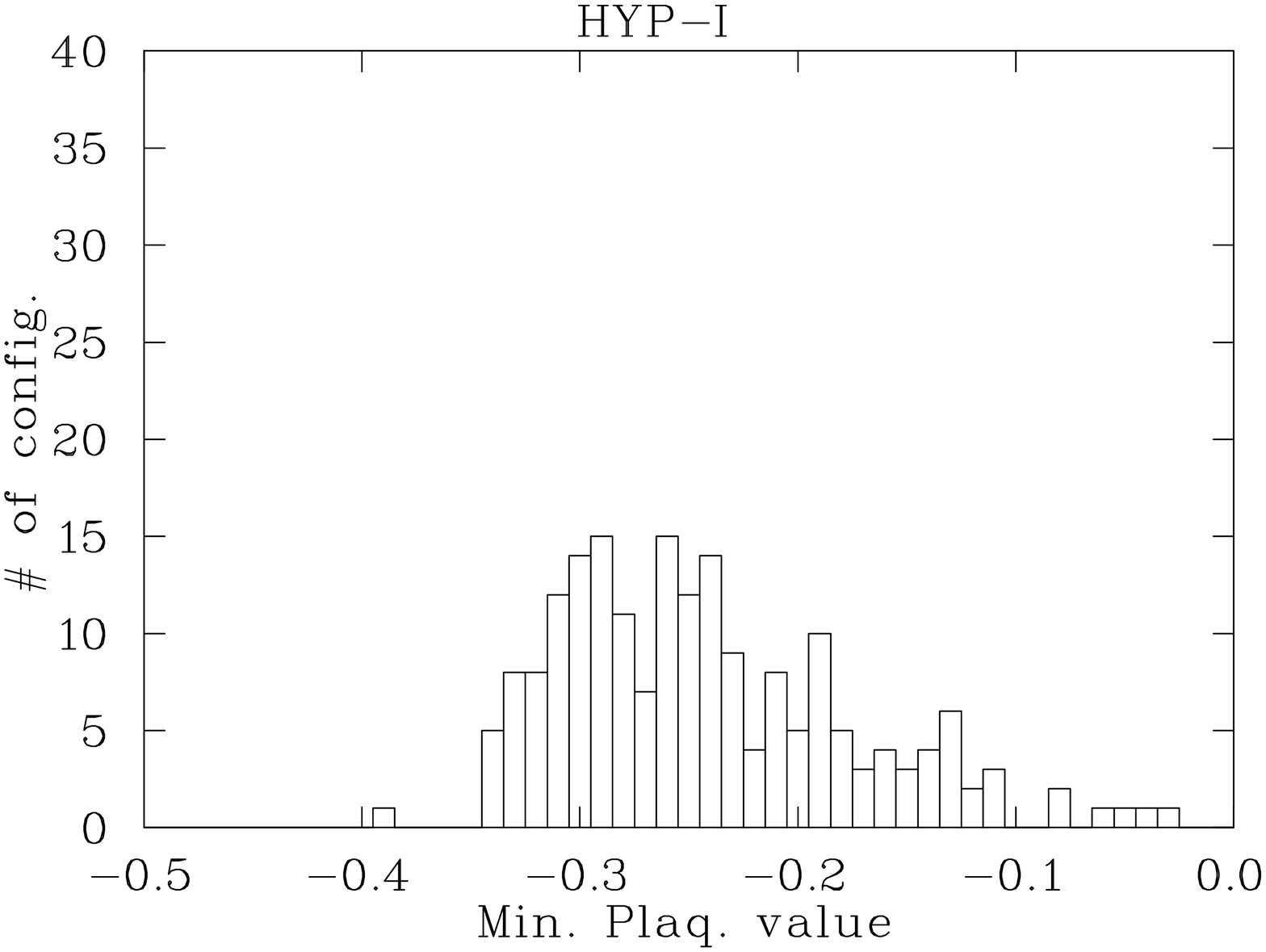} \vspace{-8mm}
\caption{Minimum plaquette values for HYP-I blocked configurations.}
\label{fig:HYP1}
\vspace{2mm}
\includegraphics[angle=90,width=15pc]{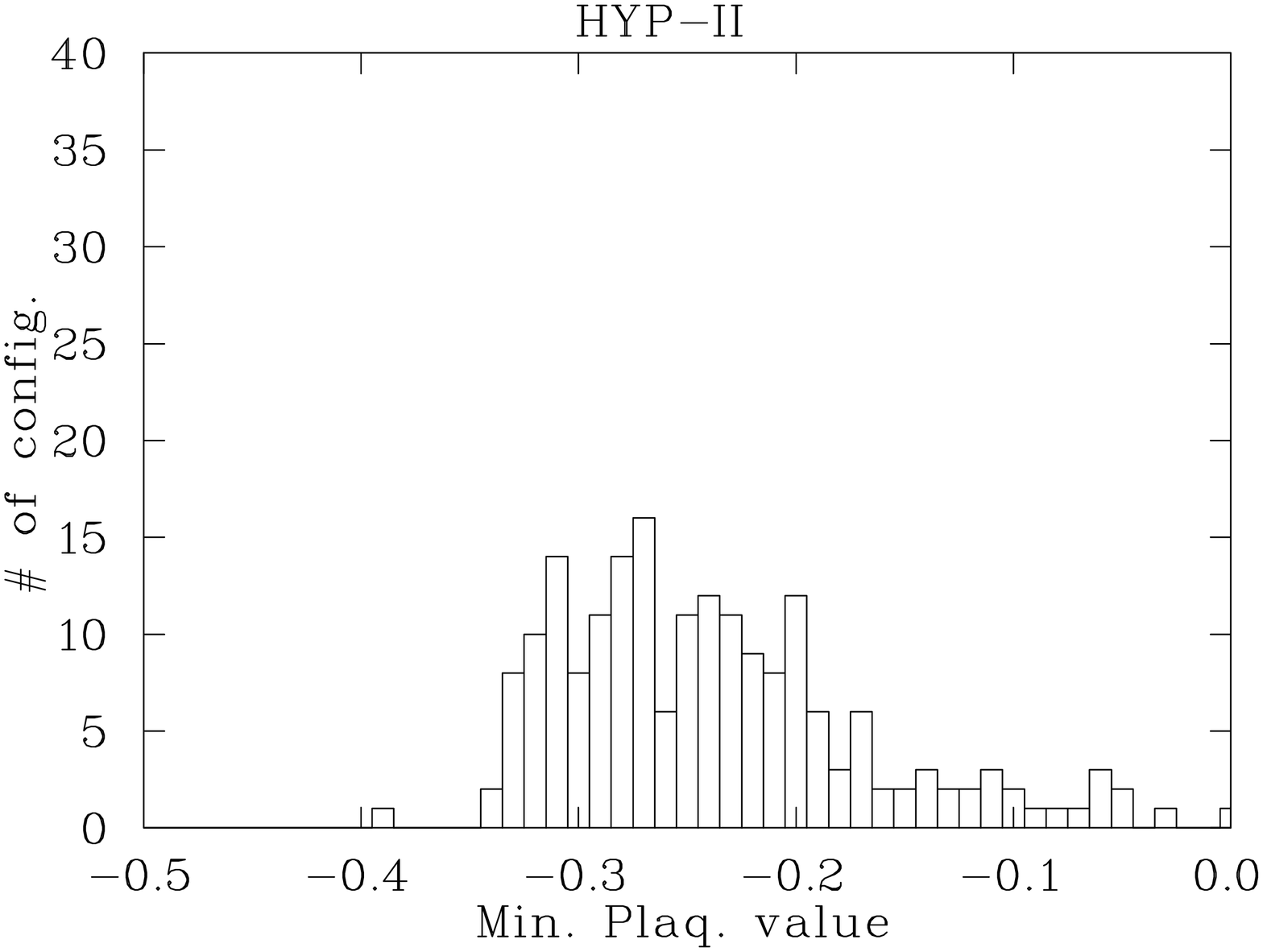} \vspace{-8mm}
\caption{Minimum plaquette values for HYP-II blocked configurations.}
\label{fig:HYP2}
\vspace{2mm}
\includegraphics[angle=90,width=15pc]{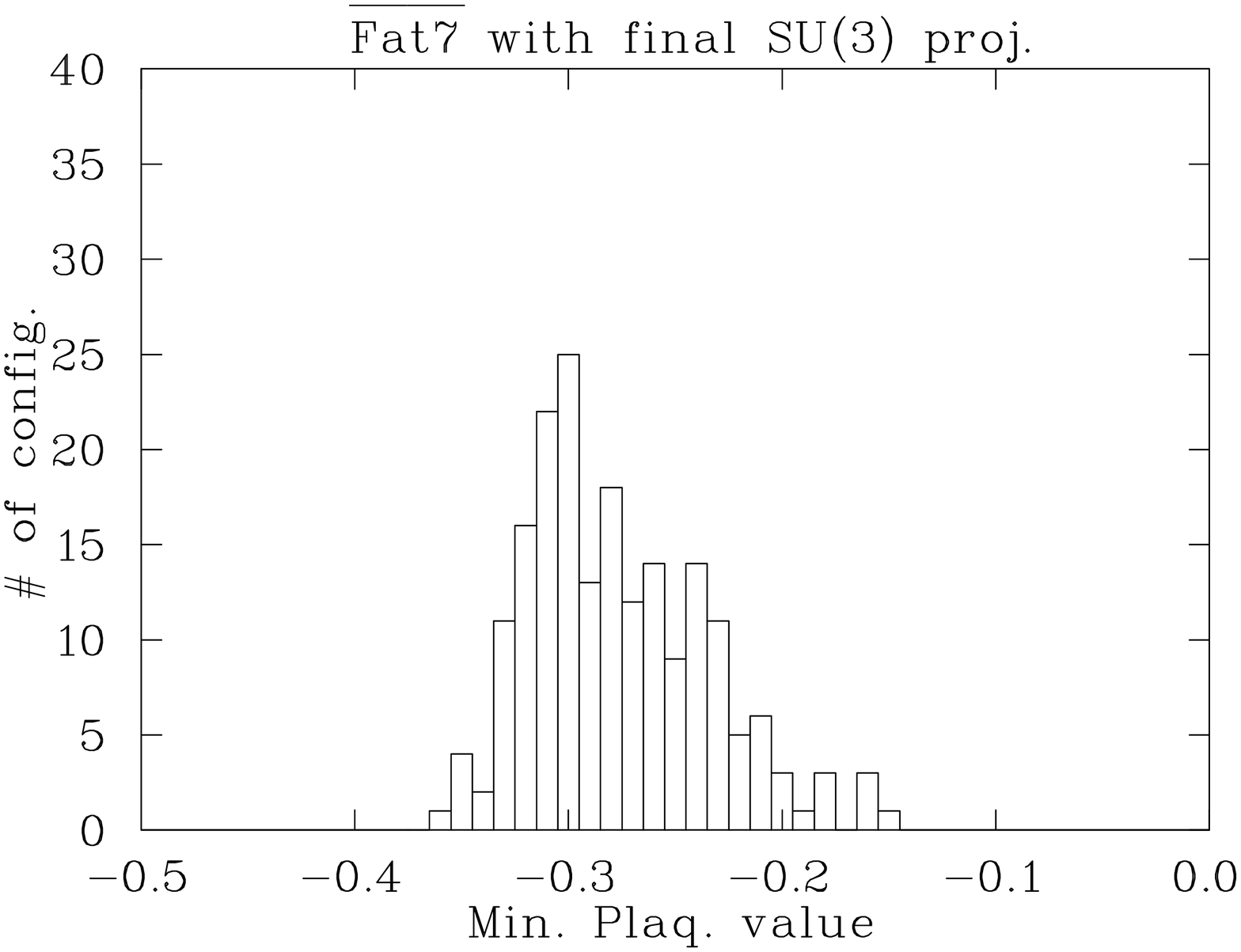}
\caption{Minimum plaquette values for $\overline{\rm Fat7}$-blocked 
configurations, with final stage SU(3) projection.}
\label{fig:Fat7_0}
\end{figure}

\begin{figure}
\includegraphics[angle=90,width=15pc]{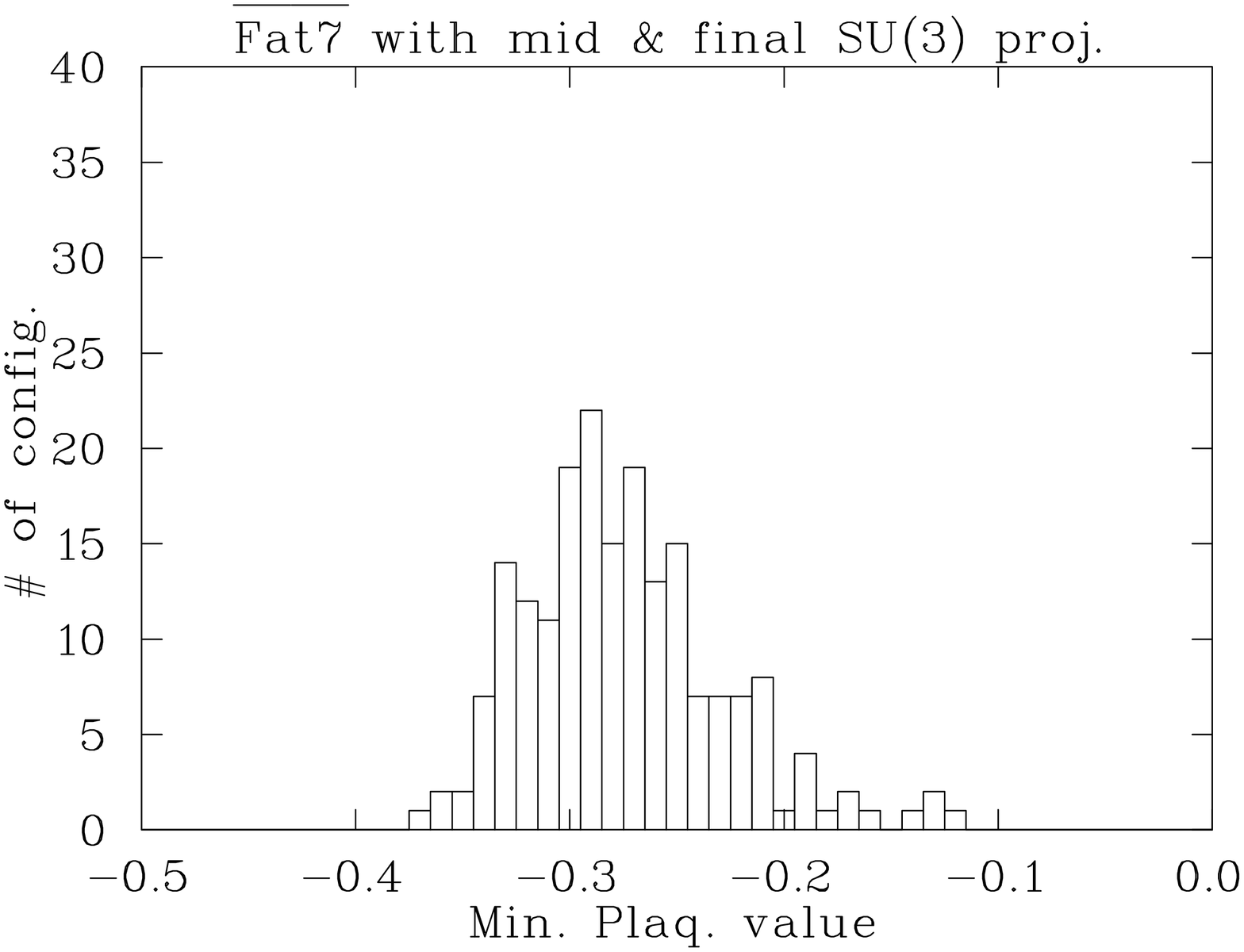} \vspace{-8mm} 
\caption{Minimum plaquette values for $\overline{\rm Fat7}$-blocked 
configurations, with middle and final stage SU(3) projection.} 
\label{fig:Fat7_I}
\vspace{2mm}
\includegraphics[angle=90,width=15pc]{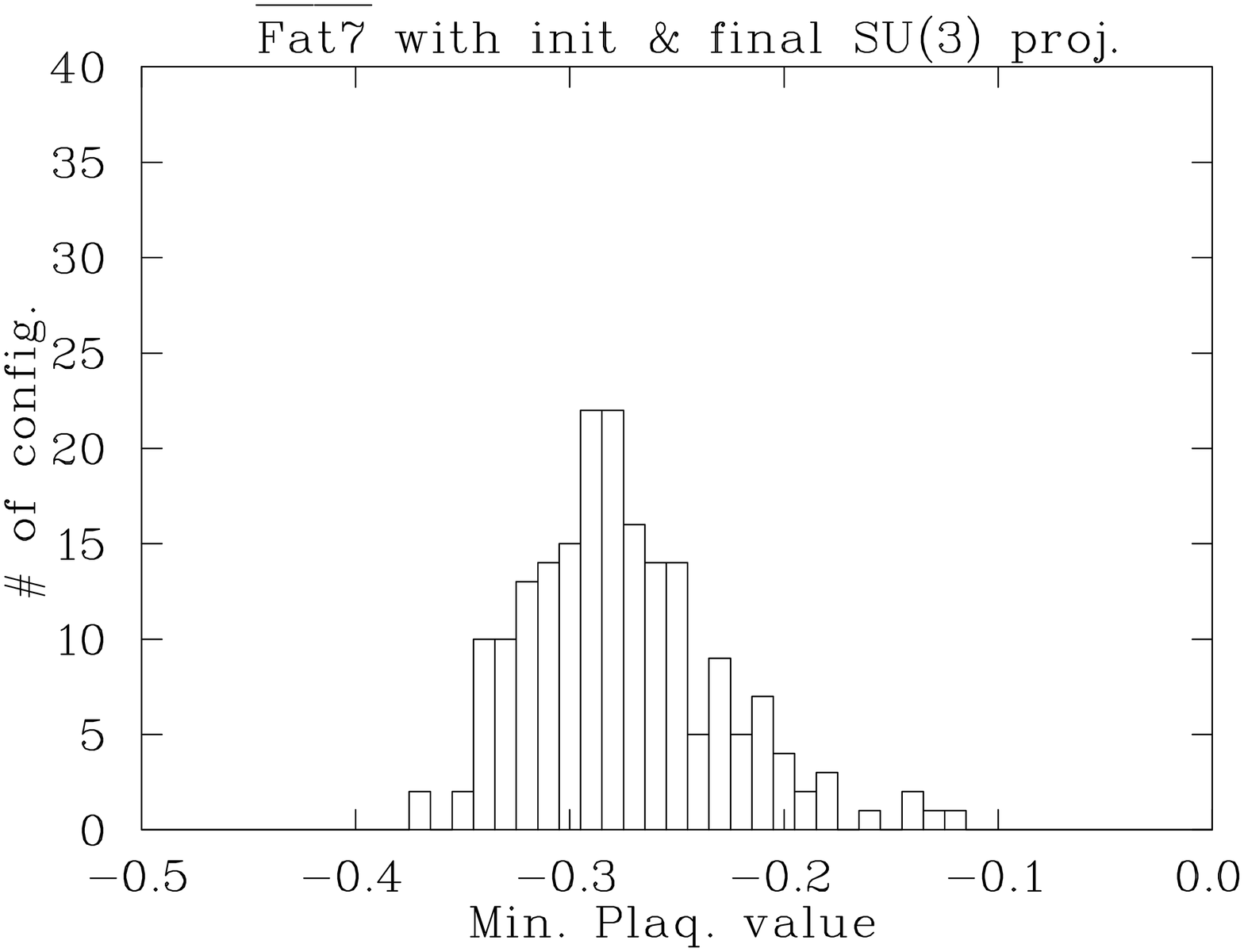} \vspace{-8mm}
\caption{Minimum plaquette values for $\overline{\rm Fat7}$-blocked 
configurations, with initial and final stage SU(3) projection.} 
\label{fig:Fat7_II}
\vspace{2mm}
\includegraphics[angle=90,width=15pc]{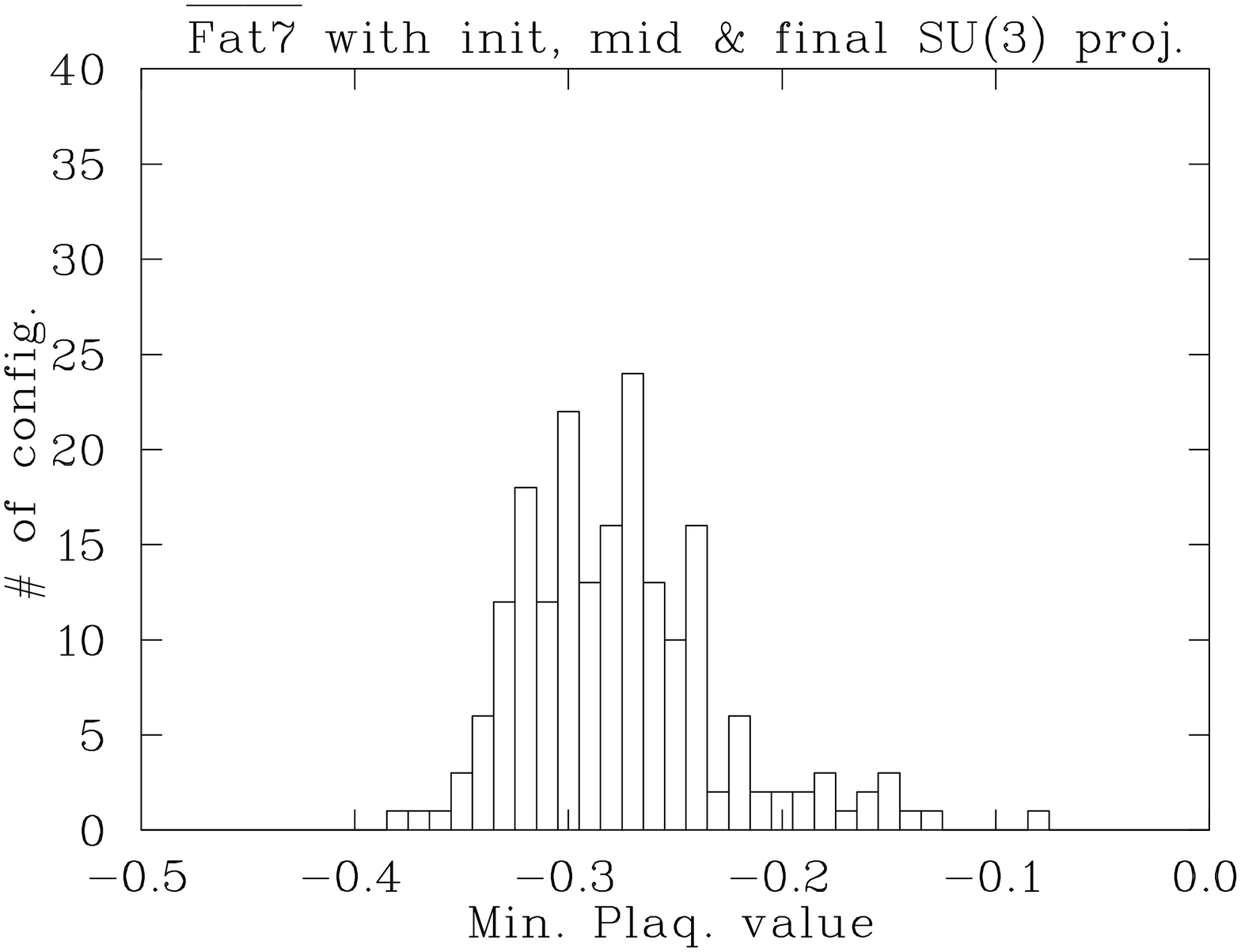} \vspace{-8mm}
\caption{Minimum plaquette values for $\overline{\rm Fat7}$-blocked 
configurations, with initial, middle, and final stage SU(3) projection.}
\label{fig:Fat7_III}
\end{figure}

We present the minimum plaquette values for an ensemble of 194 quenched 
configurations generated on an $8^3 \times 32$ lattice at $\beta = 5.7$. 
We also present the minimum plaquette values for the ensembles obtained after 
blocking each configuration with
the HYP algorithm (Figs.~\ref{fig:HYP1} and \ref{fig:HYP2}), and with the 
$\overline{\rm Fat7}$ algorithm (Figs.~\ref{fig:Fat7_0} to \ref{fig:Fat7_III}). 
As noted above the natural choice for the $\overline{\rm Fat7}$ 
weighting parameter is $\alpha=0.25$. For HYP we are interested in two choices 
of parameters, those found non-perturbatively in~\cite{hasen}, namely 
$\alpha_1=0.75$, $\alpha_2=0.6$, $\alpha_3=0.3$ (HYP-I), and those found 
perturbatively in~\cite{leesharpe}, namely $\alpha_1=7/8$, $\alpha_2=4/7$, 
$\alpha_3=1/4$ (HYP-II), although we note the caveat that choices of 
$\alpha_1$ larger than 0.75 may tend to destabilize the smearing algorithm, 
making the blocked configuration rougher rather than smoother~\cite{toomuch}. 
For the $\overline{\rm Fat7}$-blocked configurations we have analysed the 
four different SU(3) projection options described in section~\ref{sec:Fat7}. 
The histograms of minimum plaquette values are presented in 
Figs.~\ref{fig:unblock} to~\ref{fig:Fat7_III}. All figures are to the same 
scale.

We can see clearly that the initial unblocked data produce a very narrow peak. 
After HYP blocking the distribution of minimum plaquette values has been 
shifted towards (but not into) the positive region, and taken on a Poisson 
distribution-like outline. The data obtained from $\overline{\rm Fat7}$ 
with SU(3) projection at the final stage, and at all three stages are 
especially interesting from an analytical basis~\cite{wlee:0}. It is 
clear that all four $\overline{\rm Fat7}$ fattening schemes lead to minimum 
plaquette distributions that have a substantially narrower spread of values 
than the HYP-I and HYP-II data. Although the number of configurations used in 
this study is small, the results seem significant enough to warrant further 
investigations with $\overline{\rm Fat7}$.

\section{CONCLUSIONS}
Although this work is only preliminary, we find encouraging signs that 
$\overline{\rm Fat7}$ may be capable of producing results 
which are as good as HYP, and possibly better for some calculations. 
Both HYP and $\overline{\rm Fat7}$ algorithms can be made substantially faster 
and more efficient by pre-calculating and storing the staples across the entire
lattice, and using these to construct the links at the next level of fattening.
This approach avoids calculating the same staple more than once, as one would 
be forced to do if one fattened in a naive site-by-site manner. Using this 
approach $\overline{\rm Fat7}$ can be made less memory-intensive than HYP, as 
it only requires the staples in a limited number of directions to be 
pre-calculated at each stage of fattening. An efficiently-designed 
algorithm for $\overline{\rm Fat7}$ does not sacrifice speed, and enables us to 
use only one-quarter of the memory that is needed to store the 
pre-calculated staples for HYP. In future work we hope to determine whether this 
cheaper form of fattening merely equals, or in fact can produce results which 
are superior to HYP for certain calculations.

\section{ACKNOWLEDGEMENTS}
We wish to thank Anna Hasenfratz for discussions concerning the relation of 
this work to her own, and Derek Leinweber for bringing refence~\cite{toomuch} 
to our attention.


\begin{thebibliography}{8}
\bibitem{hasen} A.~Hasenfratz and F.~Knechtli, Phys. Rev. {\bf D64} (2001) 034504. 
%
\bibitem{wlee:0} W.~Lee, Phys.~Rev.~{\bf D66} (2002) 114504.
%
\bibitem{leesharpe} W. Lee and S. Sharpe, Phys. Rev.{\bf D66}, (2002) 114501. 
%
\bibitem{APE} M.~Falcioni, M.~Paciello, G.~Parisi, B.~Taglienti, Nucl. Phys. 
{\bf B251}, (1985) 624.
% 
\bibitem{giacomo} C. di Giacomo \etal, Nucl. Phys. (Proc. Suppl.) {\bf 54A},  
(1997) 343.
%
\bibitem{Fat7} K.~Orginos, R.~L.~Sugar, and D.~Toussaint, Phys.Rev. {\bf D60}, 
(1999) 054503. 
% 
\bibitem{LePage}  G.~P.~Lepage, Phys. Rev.{\bf D59}, (1999) 074502.
%
\bibitem{toomuch}  C.~Bernard and T.~DeGrand, Nucl. Phys. (Proc. Suppl.) {\bf 83}, 
(2000) 845.
%
\end{thebibliography}
\end{document}